\documentstyle[preprint,eqsecnum,aps,epsf]{revtex}	% preprint format

\newif\iftightenlines\tightenlinesfalse
\tightenlines\tightenlinestrue

\begin{document}
%
%%%%%%%%%%%%%%%%%%%%%%%%%%%%%%%%%%%%%%%%%%%%%%%%%%%%%%%%%%%%%%%%%%%%%%%%%%%%%%%
%\def\pT{p_T^{\phantom{7}}}
%\def\MW{M_W^{\phantom{7}}}
%\def\ET{E_T^{\phantom{7}}}
%
\draft
%%%%%%%%%%%%%%%%%%%% TITLE PAGE %%%%%%%%%%%%%%%%%%%%%%%%%%%%%%%%%%%%%%%%%%%%%%
\preprint{\vbox{\baselineskip=14pt%
   \rightline{FSU-HEP-951229}\break 
   \rightline{UIOWA-95-07}
}}
\title{
%\begin{title}
MULTIPLE PARTON EMISSION EFFECTS\\
IN NEXT-TO-LEADING-ORDER \\
DIRECT PHOTON PRODUCTION
%\end{title}
}
\author{Howard Baer$^1$ and Mary Hall Reno$^2$}
\address{
%\begin{instit}
$^1$Department of Physics,
Florida State University,
Tallahassee, FL 32306 USA
%\end{instit}
}
\address{
%\begin{instit}
$^2$Department of Physics and Astronomy,
University of Iowa,
Iowa City, IA 52242 USA
%\end{instit}
}
\date{\today}
\maketitle
\begin{abstract}

A recent global analysis of direct photon production at hadron collider and
fixed target experiments has noted a disturbing trend of disagreement between 
next-to-leading-order (NLO) calculations and data.
The conjecture has been made that the discrepancy is due to
explicit multiple parton emission effects which are not accounted for 
in the theoretical calculations. 
We investigate this problem by merging a NLO calculation
of direct photon production with extra multiple parton emissions via
the parton shower (PS) algorithm. 
Our calculation maintains the integrity of
the underlying NLO calculation while avoiding ambiguities due to 
double counting of multiple parton emissions. 
We find that the NLO+PS calculation
can account for much of the theory/CDF data discrepancy at $\sqrt{s}=1.8$ TeV.
It can also account for much of the theory/UA2 discrepancy if a very large
virtuality is assumed to initiate the initial state parton shower.
For lower energy data sets ({\it e.g.} $\sqrt{s}< 63$ GeV), 
NLO+PS calculations alone cannot account for the data/theory discrepancy,
so that some additional non-perturbative $k_T$ smearing is needed.
\end{abstract}

\medskip
\pacs{PACS numbers: 12.38.Bx, 12.38.Qk, 13.85.Qk, 14.80.Am}
%{\tt$\backslash$\string pacs\{\}}

\narrowtext

%%%%%%%%%%%%%%%%%% MAIN TEXT %%%%%%%%%%%%%%%%%%%%%%%%%%%%%%%%%%%%%%%%%%%%%%%

\section{Introduction}

Direct photon production\cite{halzen,cont,co,bbf,rmp} in hadronic collisions 
has long been 
recognized as an important testing ground for perturbative QCD since many 
of the ambiguities involved in measuring jets are not present when
analyzing photons. Direct photon production in lowest order QCD takes place
via annihilation and Compton scattering Feynman graphs. Since the Compton
graph involves initial state gluon-quark scattering, measurements of direct 
photon events can serve as important constraints in the determination
of the gluon
parton distribution function\cite{rmp}. 
For such a program to proceed, the greater
precision involved in next-to-leading order (NLO) QCD
calculations for the hard scattering are used. NLO calculations 
for parton+parton$\rightarrow \gamma X$ have been performed 
both analytically\cite{aurenche} and in a Monte Carlo framework\cite{boo}. 

A recent global analysis
of direct photon production in hadron collisions has noted a discrepancy
between NLO calculations and a large array of data for the transverse momentum
$p_T$ distributions of the photon\cite{cteq}.
Characteristically, in both fixed target and collider experiments, 
there is an experimental excess of photons at low transverse momentum.
Several possible explanations have been put forth to resolve the
discrepancy. These include {\it i}) improved (NLO) treatment of bremsstrahlung 
contributions\cite{gluck} and isolation criteria\cite{vv}, 
{\it ii}) modifying gluon distribution 
functions and QCD scale choices to improve the data/theory agreement\cite{vv},
or usage of alternative parton distribution functions (PDF's)\cite{quack},
and {\it iii}) invoking additional partonic $k_T$ smearing effects\cite{cteq}. 
The latter case comes in two different guises: extra partonic $k_T$ can come
from non-perturbative effects from parton binding and 
intrinsic transverse momentum, or from additional hard multiple parton 
emissions which can be calculated or modeled in perturbative QCD. 
The non-perturbative effects are generally implemented as 
Gaussian smearing in an attempt to match the data. The perturbative
multiple gluon emission effects can be implemented via even higher 
(but fixed) order perturbative calculations, via multiple gluon resummation
techniques, or via the parton shower (PS) algorithm\cite{fw,backsh}.
The resummation and PS approaches both involve approximate 
{\it all orders} perturbative QCD effects.

In this paper, we explore the extent to which the direct photon 
data can be explained by
combining a NLO QCD calculation with multiple parton emission via the 
parton shower algorithm. 
In doing so, we follow generally the prescription outlined in 
Ref.~\cite{baerreno}, where NLO $W$ and $Z$ boson
production were merged with parton showers. In these calculations, 
Owens' method of phase space slicing is used to evaluate the NLO
cross sections\cite{owens}. This method lends itself to a direct implementation
of parton showers wherein a potential problem of double counting multiple
parton emissions can be avoided. We show that our implementation
of showering with the NLO QCD calculation yields
an excess of events at
low $p_T$ relative to the unshowered NLO result
at the Fermilab Tevatron and 
CERN S$p\bar p$S energies, qualitatively accounting
for the discrepancy between theory and experiment. Additional non-perturbative
smearing is required for lower energies characteristic of 
the CERN ISR or fixed target experiments.

\section{Calculational Procedure}

Central to our calculation of direct photon production is the numerical
integration of phase space via
Monte Carlo methods\cite{boo}.
One begins by evaluating the 
${\cal O}(\alpha\alpha_s )$ and ${\cal O}(\alpha\alpha_s^2 )$ direct 
photon production subprocess Feynman graphs, including bremsstrahlung
corrections to $q\bar{q}\to q\bar{q}$, {\it etc.} 
Dimensional 
regularization is used here for ultraviolet, soft and collinear
singularities. The
four-momenta for the
$2\to 2$ subprocess are labeled according to, for instance,
$g(p_1)+q(p_2)\to \gamma (p_3)+q(p_4)$; similarly, for $2\to 3$ subprocesses,
we use $g(p_1)+q(p_2)\to \gamma (p_3)+q(p_4)+g(p_5)$, {\it etc.}
Ultraviolet singularities are renormalized using the 
$\overline{\rm MS}$ prescription\cite{aurenche}. Collinear singularities are 
factorized and then absorbed into parton distribution functions (PDF's) 
or fragmentation functions. Soft singularities are canceled between $2\to 3$
graphs and $2\to 2$ graphs. At this point, all cross section contributions
are finite, so that numerical predictions can be made. 

What is peculiar to
the Monte Carlo method of NLO calculation used here is that the phase space 
integrations are done partly analytically, and partly numerically.
The boundary between numerical and analytical methods is chosen
by selecting two theoretical cutoffs
to demarcate the collinear and soft regimes.
If any invariant quantity $t_{ij}\equiv (p_i-p_j)^2$ from the $2\to 3$
subprocess has a value $|t_{ij}|<\delta_c s_{12}$, where $s_{ij}
=(p_i+p_j)^2$, 
then one is in the 
collinear regime.
In this regime, the matrix element squared is evaluated in the leading
pole approximation and the integration 
near the collinear pole is done analytically. The cross section 
contribution is {\it de facto} $2\to 2$, and it is combined with 
the leading order and virtual 
contributions to the $2\to 2$ subprocesses. If a final state gluon energy
(in the subprocess rest frame) has value 
$E_g<\delta_s \sqrt{s_{12}}/2$, then one
is in the soft regime. 
The integrations of the squared matrix elements are performed analytically 
using the soft gluon approximation, and
combined with contributions from $2\to 2$ subprocesses. The total $2\to 2$
results, after factorization, are finite, but depend on $\delta_s$ and
$\delta_c$, such that the soft and collinear singularities are recovered
in the $\delta\to 0$ limit.
The remaining  phase space
integrations are performed via Monte Carlo. This allows easy binning of 
any desired observables and allows for the simple evaluation of
the effect of experimental cuts on the NLO prediction\cite{boo,owens}.
The $2\to 3$ contributions are all positive definite over phase space, but
are also singular as $\delta_s\to 0$ or $\delta_c\to 0$. 
The $2\to 2$ contributions 
compensate the $2\to 3$ contributions and result in a total
cross section which is independent of $\delta_s$ and
$\delta_c$ over a wide range of values\cite{boo}.
The expressions for all $2\to 2$ and
$2\to 3$ processes in direct photon production, through NLO, are compiled
in Ref. \cite{boo}. This is the starting point of our evaluation of the
transverse momentum of the direct photon using a merger of NLO QCD and parton
showers.

The PS algorithm combines the simplified collinear
dynamics, represented by the $Q^2$ evolution of parton distribution
functions and fragmentation functions,
with the exact kinematics of multiple parton emission\cite{fw,backsh}.
As implemented here, no additional weights to the integral are
included with parton showers, as the $Q^2$ evolved distribution functions
and fragmentation functions are used in evaluating the differential
cross section.
For the direct photon transverse momentum distribution,
initial rather than final state showering is most important.
Using a backward shower algorithm\cite{backsh}, 
the initial state showers are evolved
backward from a starting virtuality $t_v$.
The kinematics of the multiple partons
in the initial state shower result in transverse
momenta for the partons participating in the hard scattering,
effectively boosting the direct photon transverse momentum
relative to the collinear approximation of the kinematics.
In practice,
the parton shower is cutoff at some low $t_{\rm min}$ 
value where perturbative QCD
is still valid, but where the multiple emissions no longer become
resolvable. In all the results described below, we set $t_{\rm min}=5$
GeV$^2$.
Different prescriptions have been worked out for modeling
final state showers\cite{fw} as opposed to initial state (backward) 
showers\cite{backsh}.
At this stage, in a full simulation,
the explicit parton emissions would be combined with
a hadronization model which converts the partons into detectable particles.
Our calculation does not include hadronization.
The inclusion of hadronization should not alter our conclusion
that multiple parton emission in the initial state can qualitatively
account for
the discrepancy between theory and experiment in direct photon production.

While the PS prescription for LL calculations is straightforward, the
prescription for merging PS with NLO calculations is not. One problem is that
the shower emission from a $2\to 2$ subprocess may be double counted by 
the exact emission of an extra parton in the $2\to 3$ subprocess. 
Another problem is that, to be consistent, NLO dynamics should be 
used to govern the parton shower development. We use 
initial and final state shower algorithms consistent with LL dynamics, 
although we
use the NLO parton distribution functions in our calculation of
initial state shower probabilities.
Consequently, our calculation is not consistent to NLO: the PS 
algorithm here should be regarded only as a parametrization of a fully
consistent NLO PS program. From a practical standpoint, the error induced
by using only collinear dynamics in the PS algorithm in the first place 
should be far larger than the error induced by neglecting NLO corrections 
to the underlying collinear shower dynamics.
Our goal
here is to demonstrate that multiple parton emissions may be responsible
for the discrepancy between data and theory at low transverse momentum.

To avoid the double counting problem, we restrict shower development to the
$2\to 3$ subprocesses in
which all momentum vectors are large and well separated. 
One can view a Monte Carlo NLO 
calculation as a sort of truncated parton shower, with only a single extra 
parton emission, but which is performed exactly 
to ${\cal O}(\alpha\alpha_s^2 )$. In this case, the $2\to 2$ contributions,
which include various soft and collinear terms for which the starting 
shower virtuality would be tiny, would never shower. If the starting shower
virtuality is appropriately chosen for the $2\to 3$ subprocesses, 
then only energetic,
well-separated 3-body final states will develop a parton shower. Thus, 
the third parton of the $2\to 3$ subprocess can be viewed as the first of
the potentially multiple emissions, but which is performed using exact
instead of collinear dynamics.

In our calculation of direct photon production, we have started with the NLO
calculation of Ref.~\cite{boo} merged with the PS along the lines 
of the preceding discussion. Our computer program generates $2\to 2$
subprocesses, which frequently have negative weights, along with $2\to n$
processes, with positive definite weights, but where $n\ge 3$. Crucial to
our calculation is the stipulation of the starting virtualities for the parton
shower. 

A naive choice of starting virtuality $t_v$,
such as $|t_v|=n p_T^2(\gamma )$, (with $n\sim 1$) does not ensure
that the 3-parton final state is well separated. 
This choice leads to
large amounts of showering even for soft or collinear configurations.
One example of allowed showering with $|t_v|=n p_T^2$
is a high $p_T$ photon recoiling against two nearly
collinear partons, with $|t_{45}|>\delta_c s_{12}$ but still small.
This is a region of phase space where the $2\rightarrow 2$ and
$2\rightarrow 3$ contributions at a specific $p_T(\gamma )$ may cancel.
Since showering is implemented only in the $2\rightarrow 3$ processes
and may result in a boosted $p_T(\gamma)$ for the $2\to 3$ contribution,
the required cancellation may not occur. This introduces a dependence
on $\delta_c$ (and $\delta_s$ for other configurations) which is
unphysical. In our procedure for merging NLO with PS, 
we minimize (but never completely eliminate) the dependence
of results on variations of parameters.

To minimize the dependence of results on $\delta_s$ and $\delta_c$,
we set the starting virtuality for initial state  partons to
$|t_v|=c_v {\rm min}(|t_{ij}|,s_{ij})$ for
$i,j=1-5$, namely, the minimum of all
invariants formed by the five momenta in the $2\rightarrow 3$ process,
up to a multiplicative constant $c_v$.
With this prescription, any nearly soft or collinear emissions in the 
$2\to 3$ subprocess will result in small starting virtualities, and a small
probability to shower. 
Only energetic, well separated $2\to 3$ subprocesses will 
develop a significant parton shower in the initial state. 
The final state showers are initiated with starting virtuality
$s_{12}$. Final state showers do not change  
$p_T(\gamma )$ relative to the unshowered calculation; they can, however,
affect the number of final state photons passing the isolation cut.

\section{Calculational Results and Comparison with Data}

Direct photon production data from a variety of fixed
target and collider experiments have been tabulated as a function of 
$x_T(\gamma )={{2p_T(\gamma )}/{\sqrt{s}}}$ in two recent 
studies\cite{cteq,vv}. To compare against NLO calculations, it has 
proven convenient to plot the quantity ${({\rm Data-Theory})/{\rm Theory}}$. 
Thus, data in perfect agreement with theory would lie along the $y=0$
horizontal line. In Ref.~\cite{cteq}, a common trend amongst the 
various experimental data sets was noticed, when compared against NLO QCD.
For almost all data sets tabulated, the low $x_T(\gamma )$ range was 
underestimated by the theory (NLO QCD). In Ref. \cite{vv}, the authors were
able to improve somewhat the data {\it vs.} theory discrepancy by adjusting
independently the factorization and renormalization scales. Nevertheless,
the discrepancy between data and theory persists.

In Fig. 1{\it a}, we show ${\rm({Data-NLO})/{NLO}}$ {\it vs.} $x_T(\gamma )$
for data from the CDF experiment\cite{cdf} at the Fermilab
Tevatron, using $p\bar p$ 
collisions at $\sqrt{s}=1.8$ TeV. The data points are taken from 
Ref.~\cite{cteq}, where the NLO distributions are calculated
using the CTEQ2M PDF's\cite{pdf} evaluated at the
renormalization/factorization scale $\mu =p_T(\gamma )$. 
The large 
enhancement of data over theory can be seen below $x_T(\gamma )\sim 0.05$,
which corresponds to $p_T(\gamma )\alt 45$ GeV at the Tevatron.
Our calculation employs the same scale choices as Ref. \cite{cteq}, but updated
CTEQ3M PDF's\cite{cteq3}. 
In keeping with CDF cuts, we require the photon pseudorapidity $|\eta (\gamma )|
<0.9$, and a photon isolation cut which requires that the sum of
energy, projected transverse to the beam axis, ($E_T^i$) 
of  parton $i$ within a cone of size
$\Delta R=\sqrt{(\Delta\eta)^2+(\Delta\phi)^2}=0.7$
satisfy
$$\sum_i E_T^i\Biggr|
_{\Delta R=0.7}<2\ {\rm GeV}.$$
These two cuts are also used in Figs. 2 and 3 below.

To minimize differences due to parton distribution choices, {\it etc.},
rather than comparing the data to our NLO calculation 
merged with parton
showers (NLO$\oplus$PS), we show the effect of showering as an excess or
deficit relative to the unshowered NLO calculation.
In Figs.~1{\it b} and 1{\it c}, we show  the relative $x_T(\gamma )$ 
distributions
(NLO$\oplus$PS-NLO)/NLO where the initial state virtuality 
is chosen with $c_v=4$. In our calculation,
we have run for subprocess photon $p_T(\gamma )>4$ GeV, 
since the matrix elements are singular
as $p_T(\gamma )\to 0$; the results do not change noticeably if instead we use
$p_T(\gamma )>2$ GeV. 
Fig.~1{\it b} employs $\delta_s=10 \delta_c=0.1$,
and Fig.~1{\it c} has $\delta_s=10 \delta_c=0.02$.
We see in Figs. 1$b$ and 1$c$ that the incorporation of the PS has led
to an enhancement of the relative $x_T(\gamma )$ distributions
at $x_T(\gamma )\sim 0.02$
of about $30-40\%$, and hence is in accord with the 
data for the low range of $x_T(\gamma )$. The enhancement has been traced 
to the fact that a small fraction of the large population of very low
$x_T(\gamma )$ photons gets boosted up to higher energies by recoiling 
against the multiple parton emissions.
Although the enhancement at low $x_T(\gamma )$ from the 
NLO$\oplus$PS calculation is similar for the two cases, 
the large relative $x_T(\gamma )$ distributions show a deficit of
$10-20\%$. 
The high $x_T(\gamma )$
deficit is due to the effect of the photon isolation cut. 

For very
high energy events, there can still exist significant shower virtualities
for events with quasi-soft or collinear partons, which 
introduces a slight dependence on $\delta_s$ and $\delta_c$. There is some
enhancement in showering for very high energy events, which leads to fewer
isolated photons, and a net diminution of signal due to the isolation cut.

If we modify the initial shower virtuality magnitude by varying $c_v$,
we find that a choice of $c_v\sim 1$ results in modest enhancements of
the low $x_T(\gamma )$ region by only $\sim 10\%$. Choosing $c_v$ as high as
$c_v\simeq 9$ yields enhancements typically around $80\%$. Also, 
we have investigated how the results change by changing the initial state 
shower cutoff virtuality choice from $t_{\rm min}=5$ GeV$^2$ to 
$t_{\rm min}=3$ GeV$^2$. The latter variation yields typically a $20\%$
effect. In spite of these various uncertainties, the overall qualitative
trend of enhanced cross section at $x_T(\gamma )\alt 0.06$ persists in all
the cases we have examined. 

In Ref.~\cite{cteq}, it was noted that an ad-hoc Gaussian smearing of 
the subprocess $p_T$ led to improved agreement between theory and data. In 
Fig.~1{\it d}, we additionally introduce Gaussian smearing (GS) 
to both $2\to 2$ and $2\to 3$ processes, with average
transverse momentum zero and width $\sigma =1$ GeV.
The overall enhancement of the NLO$\oplus$PS at $x_T(\gamma )\sim 0.02$
remains, but with some slight additional enhancement 
for NLO$\oplus$PS$\oplus$GS at even lower $x_T(\gamma )$ values.
The small effect of the Gaussian smearing at CDF is not surprising since the
average boost generated by the PS algorithm is $\sim 2.5$ GeV.

In Fig. 2{\it a}, we show data from the UA2 experiment\cite{ua2} 
($p\bar p$ collisions at $\sqrt{s}=630$ GeV) compared with NLO QCD, 
for scale choice $\mu =p_T(\gamma )/2$. 
Here we use a photon $p_T$ cutoff of $p_T(\gamma )>2$ GeV.
Again, we see that
data exceeds theory by $\sim 40\%$, although this time 
for $x_T(\gamma )\sim 0.05$ (corresponding to $p_T(\gamma )\sim 16$ GeV).
In Fig. 2{\it b}, we plot the NLO$\oplus$PS 
result, using again the initial state
virtuality choice $c_v=4$, and for $\mu=p_T(\gamma)/2$ and 
$\delta_s=10\delta_c=0.02$. 
Our merged NLO$\oplus$PS calculation 
gives an enhancement of
$\sim 20\%$ above NLO results for $x_T(\gamma )\sim 0.05$.
Although the CDF and UA2
calculations start with similar virtualities, the relatively higher 
value of Feynman-$x$ in the UA2 case leads to lesser amounts of initial 
state PS radiation. This can be offset to some extent by choosing a higher 
starting virtuality, $c_v=9$, shown in Fig. 2{\it c}. The increase in 
virtuality leads to a rise in our calculation to about $40\%$ above NLO
expectations, in accord with the data. Finally, in Fig. 2{\it d}, we 
include as well the Gaussian smearing, which leads to some additional
enhancement at low $x_T(\gamma )$.

Finally, we turn to much lower energy $pp$ collider results from
experiments at the CERN ISR at $\sqrt{s}=63$ GeV.
In Fig. 3$a$, we show the data from the R806 experiment\cite{isr} 
compared with NLO
QCD for $\mu=p_T(\gamma)/2$. Using the same scale $\mu$, and including parton
showers, we show in Figs. $3b$ and $3c$, the comparison
(NLO$\oplus$PS-NLO)/NLO for $c_v=4$ and $c_v=9$, respectively. 
We have lowered the photon $p_T$ cutoff here to $p_T(\gamma )>1$ GeV.
Because of 
the large values of parton $x$ and small virtualities, at this energy,
there is very little showering, so that perturbative multiple parton
emission as described by the PS algorithm cannot explain the data/theory
discrepancy. However, Gaussian smearing on the
order of 1 GeV can be a large effect at this energy, where 
$x_T(\gamma)=0.1-0.4$ corresponds to $p_T(\gamma )=3-13$ GeV. 
In Fig. $3d$, we invoke as usual the $\sigma\sim 1$ GeV Gaussian smearing 
of the subprocess transverse momentum. In this case, the
smearing can move the low $x_T(\gamma )$ theoretical prediction into 
rough agreement with the data.

\section{Summary and Conclusions}

In summary, we have investigated the effects of multiple parton emissions
on direct photon production in hadronic collisions by merging the PS
technique with NLO QCD. For experiments at very high energy 
({\it e.g.} UA2 and CDF), the extra $k_T$ smearing of the hard scattering 
subprocess induced by the multiple parton emissions can cause some of
the relatively numerous low $p_T$ photons from NLO QCD to be boosted to higher
$p_T$ values. Such an effect causes a shift in the predicted $x_T(\gamma )$
distribution, thereby {\it improving} the agreement between theory and 
experiment. Our results cannot be interpreted as a QCD prediction
due to the many uncertainties in the PS algorithm, 
and in our merging procedure.
Amongst these uncertainties are the nature of the PS algorithm itself, and
the prescription for initial and cutoff virtualities in the PS. On the 
other hand, our results can be interpreted as an existence proof that 
higher order effects (particularly from multiple parton emission) can
account for the  theory {\it vs.} data discrepancy. Other 
groups\cite{vv,quack} have noted that the theory {\it vs.} data discrepancy
can be resolved in NLO QCD mainly by using modified parton distribution
functions. We comment 
that our result of an appropriately shifted $x_T(\gamma )$
distribution will obtain for any choice of PDF's or hard scattering scale
choices, as long as sufficient parton showering can be produced. 
Since hard scattering processes in nature are of course {\it all-orders}
processes, one would expect at some level a discrepancy between data and 
fixed order QCD to occur. Our results show that this may already be the case
for the direct photon $x_T(\gamma )$ distributions. 

For lower energy data sets ({\it e.g.} $\sqrt{s}\alt 63$ GeV), it is 
difficult to produce sufficient QCD radiation via the PS to improve the 
theory {\it vs.} data discrepancy. We do note, as in Ref. \cite{cteq}, that
an intrinsic Gaussian $k_T$ smearing with width $\sigma \sim 1$ GeV will
push the theory in the right direction to match with data. Thus, the theory
{\it vs.} data discrepancy can be resolved globally by invoking 
extra $k_T$ for the hard scattering partons: that $k_T$ would be 
primarily perturbative
in nature for high energy data sets, but mainly non-perturbative for data sets
taken at $\sqrt{s}\alt 100$ GeV.

%%%%%%%%%%%%%%%%%%%%%%%%% ACKNOWLEDGEMENTS %%%%%%%%%%%%%%%%%%%%%%%%%%%%%%%%%%%%%

%\newpage
\acknowledgments

We thank J. Huston, S. Kuhlmann and J. F. Owens for discussions.
This research was supported in part by the U.~S. Department of Energy
under grant number DE-FG-05-87ER40319 and the National Science
Foundation Grants PHY 93-07213 and PHY 95-07688.

%%%%%%%%%%%%%%%%%%%%% REFERENCES %%%%%%%%%%%%%%%%%%%%%%%%%%%%%%%%%%%%%%%%%%%%%%
%

%
\newpage

%%%%%%%%%%%%%%%%%%%%%% FIGURE CAPTIONS %%%%%%%%%%%%%%%%%%%%%%%%%%%%%%%%%%%%%%

\begin{figure}
\caption[]{$x_T(\gamma )$ distribution for $p\bar p$ collisions at 
$\sqrt{s}=1.8$ TeV. We show {\it a}) (Data-NLO)/NLO for
CDF data, {\it b}) (NLO$\oplus$PS-NLO)/NLO for
$\delta_s=10\delta_c=0.1$, 
{\it c}) (NLO$\oplus$PS-NLO)/NLO for
$\delta_s=10\delta_c=0.02$, and {\it d}) 
(NLO$\oplus$PS$\oplus$GS-NLO)/NLO for
$\delta_s=10\delta_c=0.02$. For all plots, the hard 
scattering $p_T(\gamma )\ge 4$ GeV.}
\end{figure}
\begin{figure}
\caption[]{$x_T(\gamma )$ distribution for $p\bar p$ collisions at 
$\sqrt{s}=0.63$ TeV. We show {\it a}) (Data-NLO)/NLO for
UA2 data, {\it b}) (NLO$\oplus$PS-NLO)/NLO for
$c_v=4$, 
{\it c}) (NLO$\oplus$PS-NLO)/NLO for
$c_v=9$, and {\it d}) 
(NLO$\oplus$PS$\oplus$GS-NLO)/NLO for
$c_v=9$. For all plots, the hard 
scattering $p_T(\gamma )\ge 2$ GeV and $\delta_s=10\delta_c=0.02$.}
\end{figure}
\begin{figure}
\caption[]{$x_T(\gamma )$ distribution for $p p$ collisions at 
$\sqrt{s}=63$ GeV. We show {\it a}) (Data-NLO)/NLO for
R806 data at the CERN ISR, {\it b}) (NLO$\oplus$PS-NLO)/NLO for
$c_v=4$, 
{\it c}) (NLO$\oplus$PS-NLO)/NLO for
$c_v=9$, and {\it d}) 
(NLO$\oplus$PS$\oplus$GS-NLO)/NLO for
$c_v=9$. For all plots, the hard 
scattering $p_T(\gamma )\ge 1$ GeV and $\delta_s=10\delta_c =0.1$.}
\end{figure}
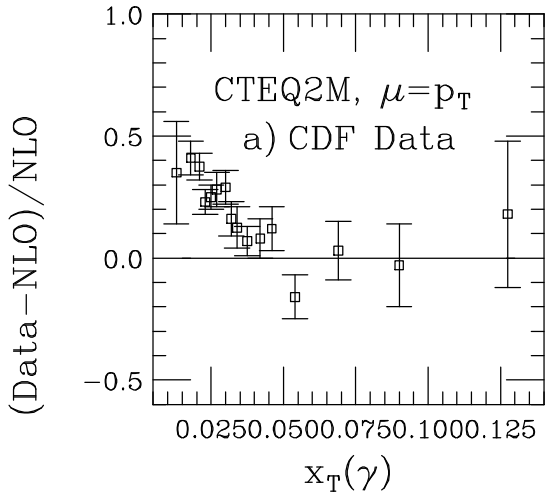
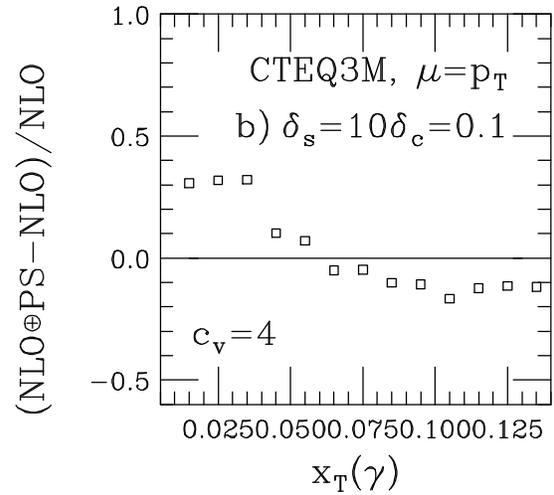
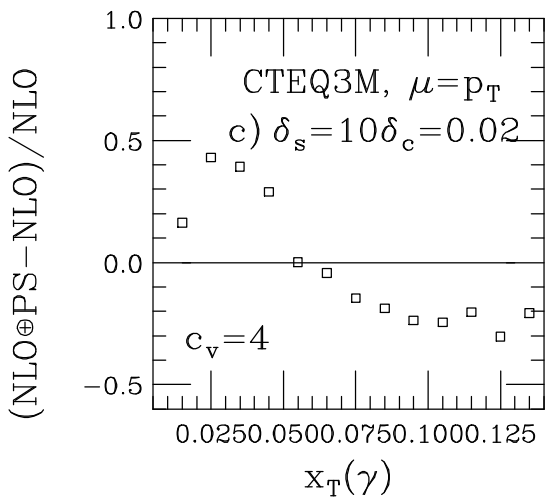
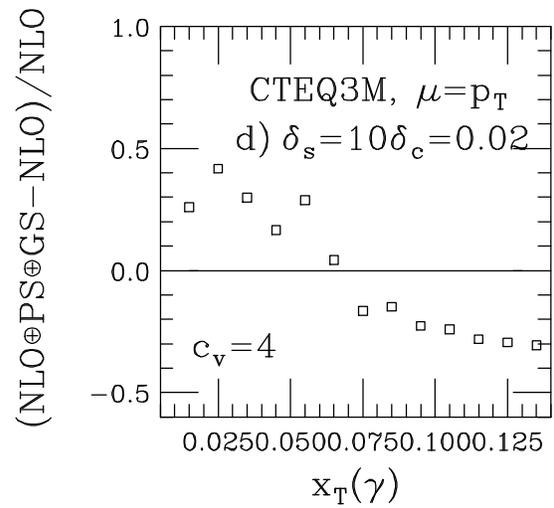
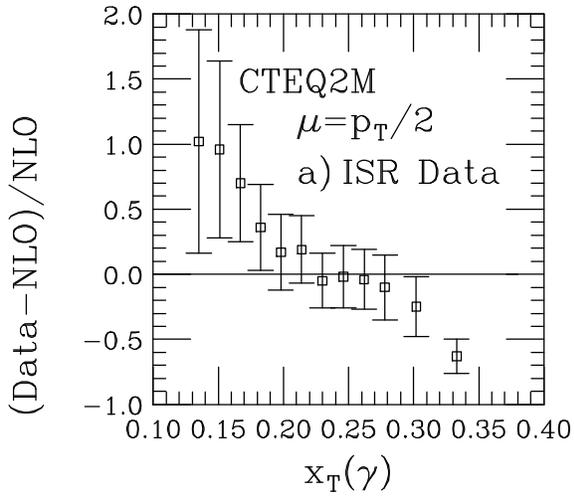
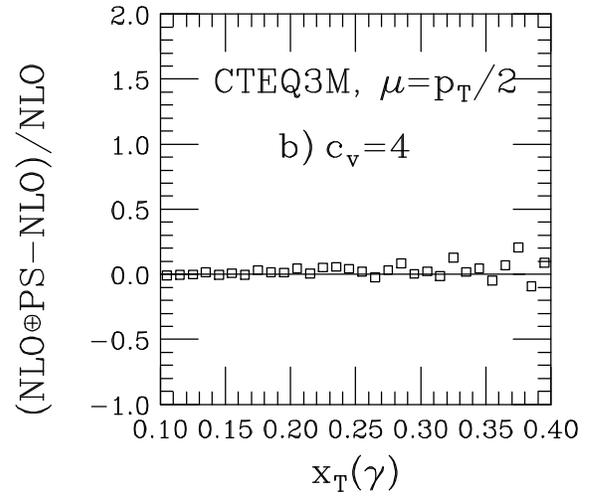
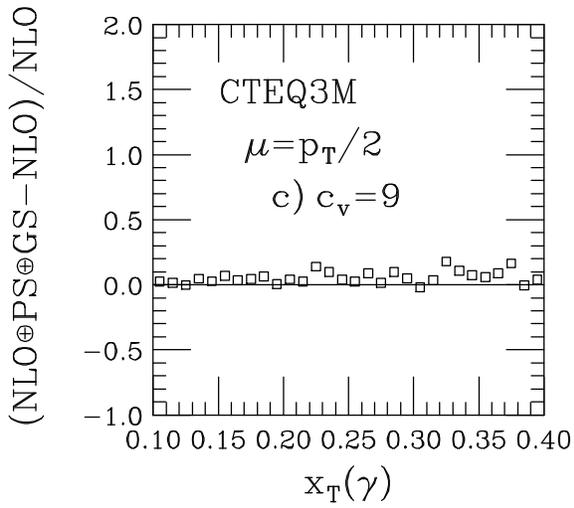
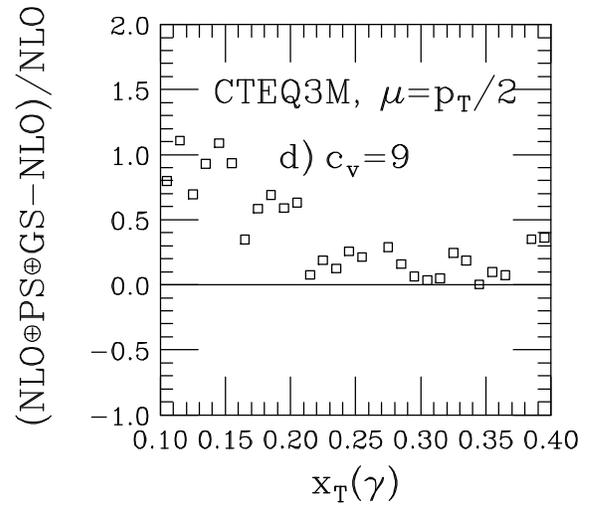
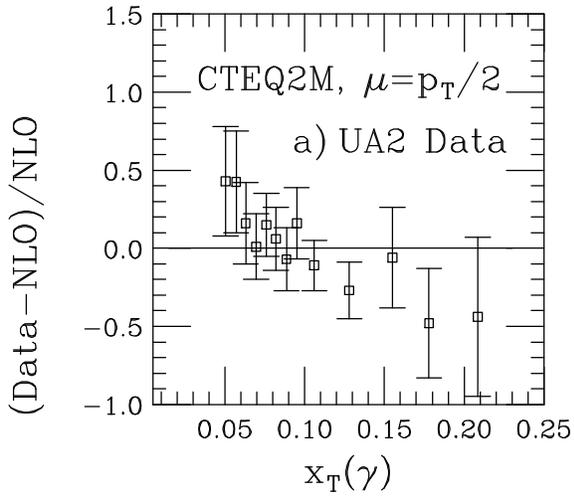
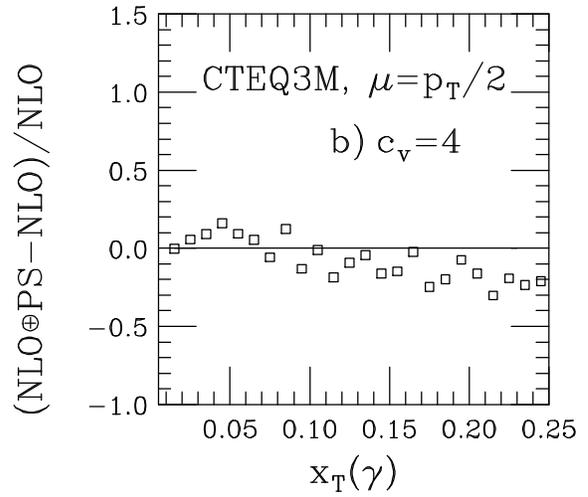
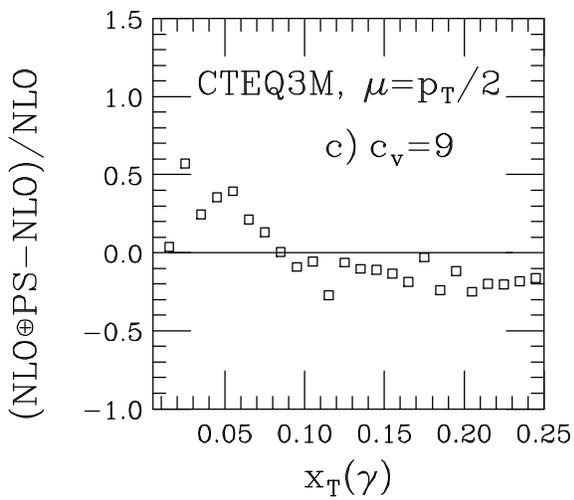
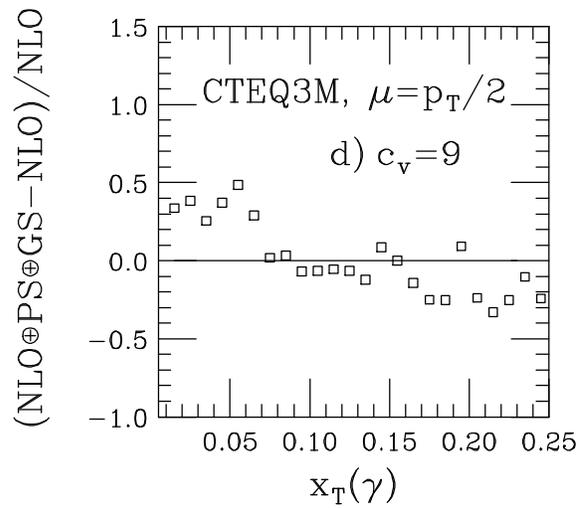
\end{document}